\documentclass[doublecol]{epl2}

\pdfoutput=1

\title{Non-stoichometry and the magnetic structure of \SCOFA}
\shorttitle{Non-stoichometry and the magnetic structure of \SCOFA}

\author{Marcus Tegel\inst{1} \and Franziska Hummel\inst{1} \and Yixi Su\inst{2} \and Tapan Chatterji\inst{3} \and Michela Brunelli\inst{4} \and Dirk Johrendt\inst{1}}
\shortauthor{M. Tegel \etal}

\institute{
  \inst{1} Department Chemie, Ludwig-Maximilians-Universit\"{a}t M\"{u}nchen, Butenandtstra\ss e 5-13 (Haus D), 81377 M\"{u}nchen, Germany\\
  \inst{2}J\"{u}lich Centre for Neutron Science, IFF, Forschungszentrum J\"{u}lich, Outstation at FRM II, Lichtenbergstra\ss e 1, D-85747 Garching, Germany\\
  \inst{3}J\"{u}lich Centre for Neutron Science, Forschungszentrum J\"{u}lich, Outstation at Institut Laue-Langevin, BP 156, 38042 Grenoble Cedex 9, France\\
  \inst{4}Institut Laue-Langevin, BP 156, 38042 Grenoble Cedex 9, France
\\\begin{center}Dedicated to Dr. Klaus R\"{o}mer on the occasion of his 70th birthday\end{center}}

\abstract{
The iron arsenide \SCOFA~with the tetragonal \SGOCS-type structure was synthesized and its crystal structure re-determined by neutron powder diffraction. In contrast to previous X-ray crystallographic studies, a mixed occupancy of chromium and iron was found within the FeAs$_{4/4}$ layer ($93\pm1$\%~Fe~:~$7\pm1$\%~Cr). We suggest that the partial Cr-doping at the Fe site is the reason for the absence of a spin-density wave anomaly and superconductivity in this compound. Additional experiments via neutron polarization analysis revealed short-range spin correlations below $\sim 100$~K and long-range antiferromagnetic ordering below $T_N = 36$~K with a magnetic propagation vector of \QVEC. The \CR~ions form a collinear magnetic structure of the $C$-type in the magnetic space group \MSG~($\mathbf{a'}=\mathbf{a-b}, \mathbf{b'}=\mathbf{a+b}, \mathbf{c'}=\mathbf{c}$), where \CR-ions occupy the $4g$ ($0,\frac{1}{4},z$) Wyckoff position. The magnetic moments are aligned along the orthorhombic $\mathbf{a'}$-axis. At 3.5~K, an ordered magnetic moment of $2.75\pm0.05~\mu_B$ for the \CR-sublattice was refined.}

\begin{document}

\newcommand{\QVEC}{$\mathbf{q} = (\frac{1}{2},\frac{1}{2},0)$}
\newcommand{\CR}{Cr$^{3+}$}
\newcommand{\TT}{$^{\circ}~2\theta$}
\newcommand{\SCOFA}{Sr$_{2}$CrO$_{3}$FeAs}
\newcommand{\SGOCS}{Sr$_{2}$GaO$_{3}$CuS}
\newcommand{\BaFA}{BaFe$_{2}$As$_{2}$}
\newcommand{\CSG}{$P\frac{4}{n}mm$}
\newcommand{\MSG}{$C_Pmma'$}
\newcommand{\CRdoped}{BaFe$_{2-x}$Cr$_{x}$As$_{2}$}
\newcommand{\PLAW}{$a * (\frac{T_N-T}{T_N})^\beta$}
\maketitle

\section{Introduction}

Since the discovery of superconductivity in tetragonal layered iron arsenides with ZrCuSiAs-type, ThCr$_2$Si$_2$-, or PbFCl-type structures \cite{Hosono-2008, Rotter-2-2008, Wang-2008} and critical temperatures up to 55~K \cite{Ren-55K, Angew-2008}, immense progress has been made regarding the rich physical and structural phenomena occurring in this new class of superconductors.\cite{Hosono-Rev, Physica-Special}. But beyond great efforts to understand the underlying physics, the search for new compounds with similar FeAs layers is also important in order to widen the material basis and perhaps to increase the critical temperatures.

\begin{figure}[h!]
\center{
\includegraphics[height=100mm]{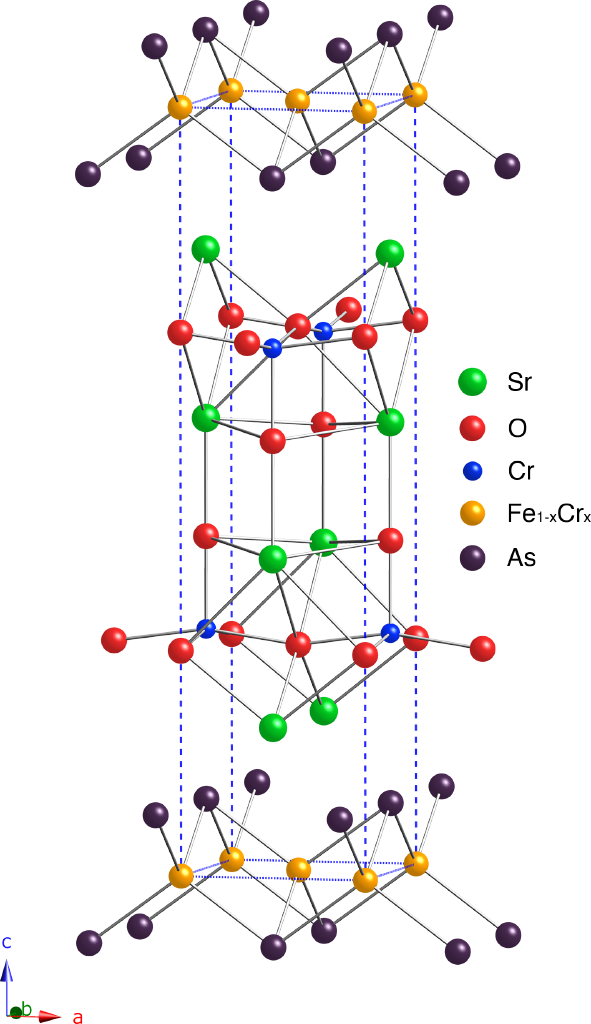}
\caption{Crystal structure of \SCOFA. Space group: \CSG, origin choice 1.}
\label{fig:Structure}
}
\end{figure}

The $T_c$'s of the known iron arsenides increase with the anisotropy of their crystal structures, which is rather small \cite{Ni+Canfield-2008} compared with the cuprates, whose $T_c$'s are also higher. Even though we should be very careful in transferring principles from the cuprates, FeAs-compounds with larger inter-layer distances are of particular interest and recently a number of new compounds with structures derived from copper sulfides with perowskite-like blocks were synthesized and studied. The first was Sr$_3$Sc$_2$O$_5$Fe$_2$As$_2$ \cite{32522-Zhu-2009} with the known structure of Sr$_3$Fe$_2$O$_5$Cu$_2$S$_2$ \cite{Hor-1997}. The iron arsenide is not superconducting and shows neither a structural anomaly nor magnetic ordering as found in the ZrCuSiAs- and ThCr$_2$Si$_2$-type parent compounds \cite{Cruz-2008, Rotter-1-2008}. Superconductivity at 17~K has been discovered in the iron phosphide Sr$_2$ScO$_3$FeP \cite{Ogino-1-2009} with the \SGOCS-type structure \cite{Hor-1997-2}. This $T_c$ is considerably higher in comparison with the ZrCuSiAs-type phosphide oxides like LaFePO (4-7~K) and may promise even higher values in the analogous arsenides. Indeed, the arsenide Sr$_2$VO$_3$FeAs with $T_c$ = 37~K was found \cite{V-21311} which proved the potential of such compounds. The isotypic chromium compound \SCOFA~(Figure~\ref{fig:Structure})\cite{Ogino-2-2009, Tegel-21311} is not superconducting, but exhibits antiferromagnetic ordering of the \CR~moments according to susceptibility measurements, whereas $^{57}$Fe-M\"ossbauer spectra revealed non-magnetic iron atoms in this compound\cite{Tegel-21311, Tegel-MB}.

The synthesis of single phase samples of the '21311'-compounds has turned out to be difficult and most of the published X-ray powder patterns reveal significant amounts of impurity phases. This is especially true in the case of the superconducting compound Sr$_2$VO$_3$FeAs and very pronounced in the recently reported Sr$_2$(Mg,Ti)O$_3$FeAs ($T_c$ = 39~K) \cite{Ogino-3-2009}, which is hardly the main phase of the sample. Such multi-phase samples cast serious doubts about the true chemical composition of the superconducting fractions.

We have synthesized almost single phase samples of the chromium 21311-compound \SCOFA, which allow a more precise determination of the structure. By neutron scattering we are able to distinguish between chromium and iron very well in contrast to X-ray diffraction. In this letter we report the re-determination of the crystal structure and the antiferromagnetic spin structure of \SCOFA. Our results  shed light on the absence of superconductivity in \SCOFA~and the chemical nature of the 21311-type compounds containing similar $d$-metals in general.

\section{Experimental}
%

\SCOFA~was synthesized by heating stoichiometric mixtures of strontium, chromium, iron~oxide and arsenic~oxide in alumina crucibles sealed in silica ampoules under an atmosphere of purified argon in four separate batches of 1 gram. Each mixture was heated to 1173~K at a rate of 80~K/h, kept at this temperature for 60~h and cooled down to room temperature. The products were homogenized in an agate mortar, pressed into pellets and sintered at 1323~K for 60 h. The batches were then united, reground, pressed into pellets of 14~mm in diameter and sintered together at 1323~K for 50~h. The obtained black crystalline product \SCOFA~is stable in air.

Powder diffraction patterns at various temperatures were recorded at the high flux powder diffractometer D20 at Institut Laue-Langevin (Grenoble, France) with  0.187~nm incident wavelength. Rietveld refinements of the D20 nuclear scattering patterns were performed with the TOPAS package\cite{TOPAS} using the fundamental parameter approach as reflection profiles (convolution of appropriate source emission profiles with axial instrument contributions as well as crystallite microstructure effects). In order to describe small peak half width and shape anisotropy effects, the approach of Le Bail and Jouanneaux\cite{LBJ} was implemented into the TOPAS program and the according parameters were allowed to refine freely. Preferred orientation of the crystallites was described with the March Dollase function. The Fe:Cr ratio on both the iron and the chromium site were also allowed to refine freely. Additionally, powder diffraction patterns were recorded using polarized neutrons (0.474~nm incident wavelength) at the polarized spectrometer DNS at FRM II (Garching, Germany) at different temperatures. An unambiguous separation of nuclear coherent, spin incoherent and magnetic scattering contributions simultaneously over a wide scattering angle has been achieved via neutron polarization analysis from the well established xyz-method\cite{xyz-method}. Both the nuclear and magnetic DNS powder patterns were interpolated, scaled and converted to a format with a constant step width and refined with the GSAS package\cite{GSAS}. The standard Gaussian profile function with asymmetry corrections (CW profile function 1) was used as reflection profiles. To obtain accurate changes of the ordered magnetic moments at different temperatures, only the gaussian parameters U and W were refined using the  3.5~K pattern and held constant for all other temperatures. The scaling factor obtained from the nuclear scattering contribution was corrected for the number of formula units per unit cell and taken as reference for the magnetic scattering contribution. A shifted Chebychev series of $9th$ order was used as background function and the parameters were allowed to refine freely for each temperature. The structural parameters for all refinements were taken from an additional refinement (combined nuclear and magnetic scattering) performed with GSAS using the low temperature diffraction data taken at D20.

\section{Results and Discussion}
\begin{figure}[h!]
\center{
\includegraphics[width=88mm]{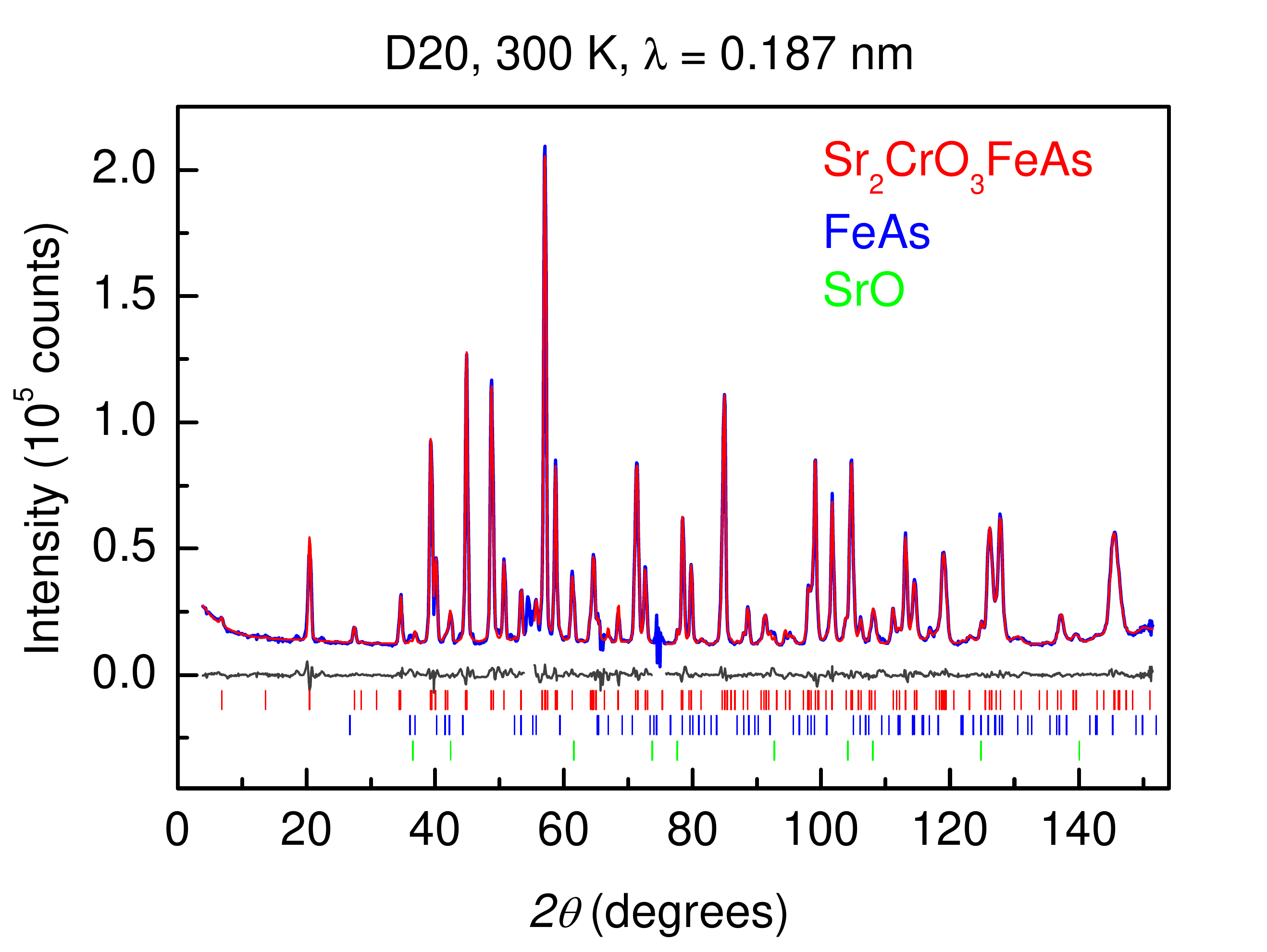}
\caption{(Color online) D20 neutron powder pattern (blue) and Rietveld fit (red) of \SCOFA~at 300~K (space group \CSG). The angle range from 54.0 to 55.1\TT~was excluded from the refinement due to the presence of an unknown impurity phase and the angle range from 73.9 to 75.7\TT~due to a faulty detector segment.}
\label{fig:Rietveld_d20}
}
\end{figure}

Figure~\ref{fig:Rietveld_d20} shows the neutron powder pattern of \SCOFA\ measured at the D20 diffractometer. It could be fitted successfully with a tetragonal \SGOCS-type \SCOFA~main phase and FeAs, as well as SrO as minor impurity phases. Yet another small impurity phase could not be identified and its largest peak was therefore excluded from the refinement. No structural phase transition was observed down to 6.5~K. The crystallographic data at 300~K (Table~\ref{tab:Crystallographic}) are in good agreement with our previously published X-ray data\cite{Tegel-21311}. However, the refinement of the neutron data unambiguously shows a mixed occupancy of iron and chromium at the iron site $2a$ ($93\pm1$\%~Fe~:~$7\pm1$\%~Cr), while no mixed occupancy at the chromium site or any oxygen deficiency were detected within one standard deviation. Within one standard deviation, full occupancy of all sites was found in X-ray diffraction experiments. These findings suggest partial interchangeability of the $3d$-metals Fe and Cr in the FeAs layers. This Cr-doping of the Fe-site could explain why \SCOFA~does neither display a spin-density-wave anomaly nor superconductivity. The situation is very similar to that in Cr-doped BaFe$_2$As$_2$, where small amounts of chromium at the iron site in \CRdoped~strongly effect the SDW anomaly and it is apparently detrimental to superconductivity \cite{Cr-doped}.

\begin{table}[p]
\caption{Crystallographic data of \SCOFA~(D20).}
\label{tab:Crystallographic}
\begin{center}
\begin{tabular}{ll}
temperature (K) & 300\\
wave length (nm) & 0.187\\
space group & \CSG~(o1)\\
 \textit{a} (pm) & 391.71(7)\\
 \textit{b} (pm) & $=a$\\
 \textit{c} (pm) & 1578.0(2)\\
 \textit{V} (nm$^{3}$) & 0.2421(1)\\
 \textit{Z} & 2\\
 data points & 1446\\
 excluded regions (\TT) & 54.0-55.1, 73.9-75.7\\
 reflections (main phase) & 118 (1 excluded)  \\
 profile variables (main phase) & 6\\
 anisotropy variables & 24\\
 atomic variables (main phase) & 15\\
 background variables & 12\\
 variables of impurity phases & 21\\
 other variables & 6\\
 \textit {d} range & $0.966 - 15.780$\\
 R$_P$, \textit{w}R$_P$ & 0.0304, 0.0452\\
 R$_{bragg}$ & 0.0110\\
 wght. Durbin-Watson $d$ stat. & 0.812\\
           &\\
Atomic parameters: \\
 Sr1 & 2$c$ ($0,\frac{1}{2},z$)\\
    & $z$ = 0.8059(3)\\
    & $U_{iso} = 131(9)$\\
 Sr2 & 2$c$ ($0,\frac{1}{2},z$)\\
    & $z$ = 0.5856(3)\\
    & $U_{iso} = 145(12)$\\
 Cr/Fe - occ. 0.99(1):0.01(1)& 2$c$ ($0,\frac{1}{2},z$)\\
    & $z$ = 0.3104(4)\\
    & $U_{iso} = 69(19)$\\
 Fe/Cr  - occ. 0.93(1):0.07(1) & 2$a$ ($0,0,0$) \\
    & $U_{iso} = 85(9)$  \\
 As & 2$c$ ($0,\frac{1}{2},z$)\\
    & $z$ = 0.0883(2)\\
    & $U_{iso} = 192(13)$\\
 O1 & 4$f$ ($0,0,z$)\\
    & $z$ = 0.2943(2)\\
    & $U_{iso} = 93(7)$\\
 O2 & 2$c$ ($0,\frac{1}{2},z$)\\
    & $z$ = 0.4308(3)\\
    & $U_{iso} = 172(12)$\\
           &\\
 \end{tabular}
 \begin{tabular}{ll}
Sel. bond lengths (pm):&\\
Sr--O  &  244.3(6)$\times$1; 251.7(3)$\times$4\\
            & 272.6(4)$\times$4; 278.2(1)$\times$4\\
Cr--O  &  190.0(8)$\times$1; 197.5(1)$\times$4 \\
Fe--Fe  &  277.0(1)$\times$4 \\
Fe--As  &  240.4(3)$\times$4\\
           &\\
Sel.  bond angles (deg):\\
As--Fe--As &  109.6(1)$\times$4; 109.1(2)$\times$2\\
O--Cr--O & 89.0(1)$\times$4; 97.4(2)$\times$4;\\
           &  165.2(4)$\times$2   \\

\end{tabular}
\end{center}
\end{table}

Such mixing of iron with other $d$-metals in the FeAs layers may also occur in other 21331 or 32522-systems, and we do not rule out that the alleged stoichiometric 37~K superconductor Sr$_2$VO$_3$FeAs\cite{V-21311} is in fact a doped compound likewise. Our data do not show any oxygen deficiencies, unlike Sr$_2$VO$_{3-\delta}FeAs$, which was recently reported\cite{V-21311-delta}. However, no detailed structural data of these compounds were published.\\

\begin{figure}[h!]
\center{
\includegraphics[height=103.5mm]{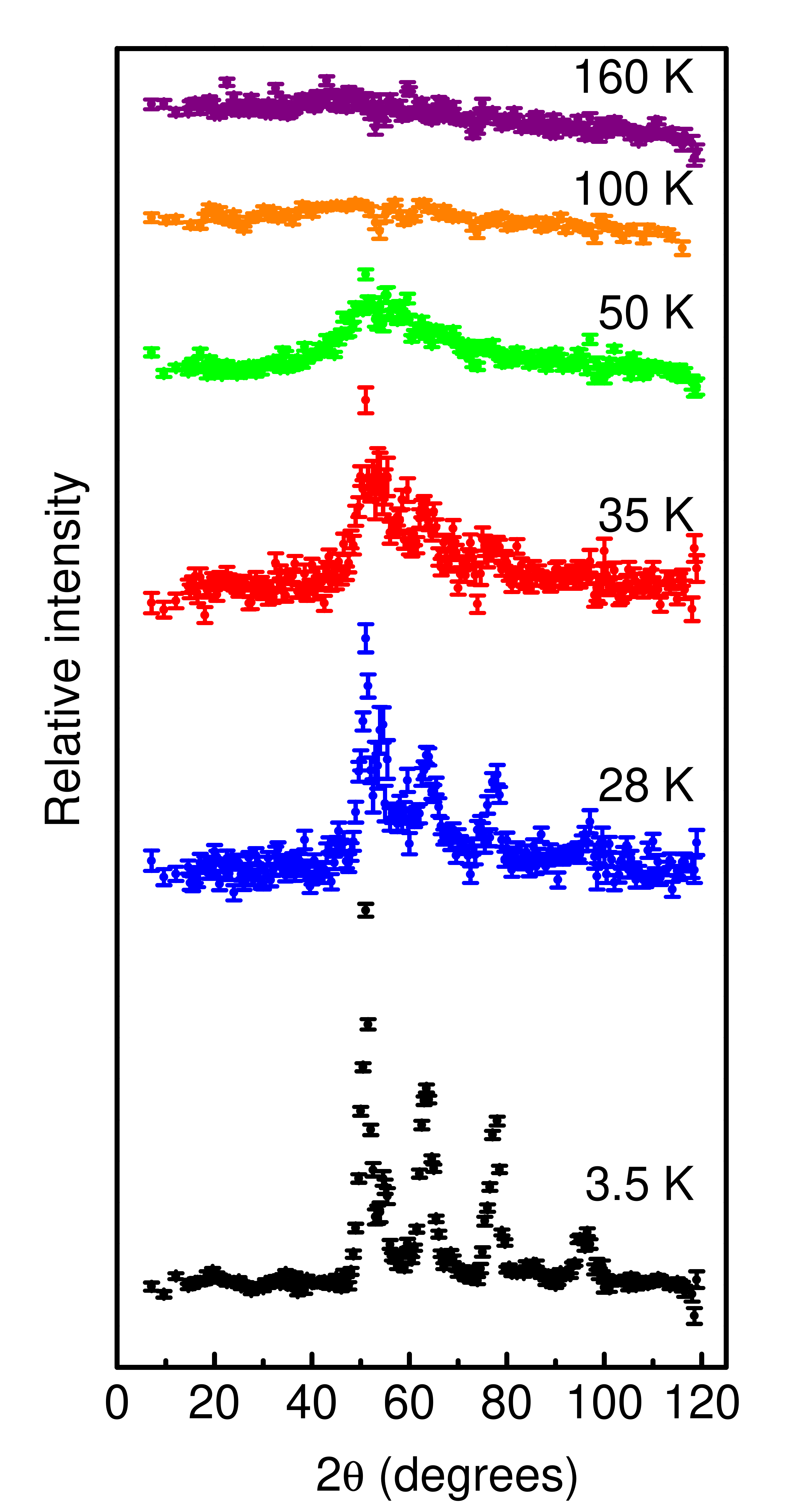}
\caption{(Color online) Evolution of the magnetic scattering contribution of \SCOFA~at different temperatures.}
\label{fig:magneticscatter}
}
\end{figure}

Recently reported susceptibility measurements on \SCOFA~\cite{Tegel-21311} revealed Curie-Weiss behavior above 150~K with an effective magnetic moment $\mu_{eff}^{exp}$ = 3.83(3)~$\mu_B$. As this is typical for \CR~ions in the $^4F_{3/2}$ state ($\mu_{eff}^{calc}$ = 3.87~$\mu_B$), we expected that the observed magnetism comes from the chromium atoms only, whereas the iron sites carry no magnetic moments. Furthermore, a drop of the $\chi(T)$ plot below 31~K together with a large negative Weiss-constant $\Theta = -141(3)$~K indicated antiferromagnetic ordering.
Neutron patterns measured at the D20 diffractometer showed additional peaks appearing below 35~K (not shown). These magnetic reflections could be indexed with the primitive tetragonal cell $|\mathbf{a'}|=|\mathbf{a}-\mathbf{b}|=\sqrt{2}\cdot |\mathbf{a}|$, $|\mathbf{c'}|=|\mathbf{c}|$ according to a magnetic propagation vector \QVEC~based on the original tetragonal unit cell.
In order to separate nuclear and magnetic scattering, we performed experiments with polarized neutrons at DNS. Sharp magnetic reflections are discernible below $T_N$, indicating long-range antiferromagnetic ordering of the Cr-sublattice (Figure~\ref{fig:magneticscatter}). 
Furthermore, magnetic diffuse scattering due to short-range spin correlations can be clearly observed above $T_N$. At temperatures above 120~K, \SCOFA~displays Curie-Weiss-like paramagnetic scattering only. Short-range antiferromagnetic spin correlations begin to emerge below $\sim 100~K$. The observed asymmetric diffuse scattering profile strongly suggests that the short-range spin correlations are two-dimensional in nature.

\begin{figure}[h!]
\center{
\includegraphics[width=41mm]{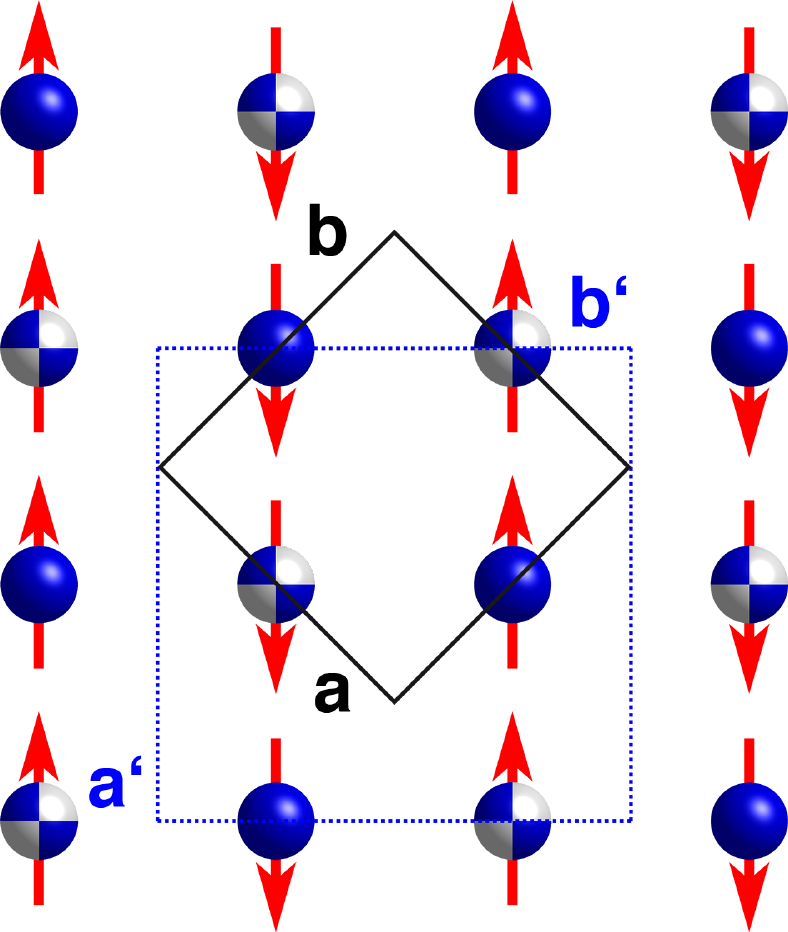}
\caption{(Color online) Magnetic ordering of the Cr atoms. Cr atoms at height $z$ are depicted as solid, Cr atoms at height $\overline{z}$ as checkered spheres. The tetragonal crystallographic unit cell is depicted as solid black, the orthorhombic magnetic cell as dashed blue square. The transformation from the tetragonal to the orthorhombic cell is $\mathbf{a'}=\mathbf{a-b}, \mathbf{b'}=\mathbf{a+b}, \mathbf{c'}=\mathbf{c}$ with an origin shift of $-\frac{1}{4},\frac{1}{4},0$.}
\label{fig:order}
}
\end{figure}

\begin{figure}[h!]
\center{
\includegraphics[height=100mm]{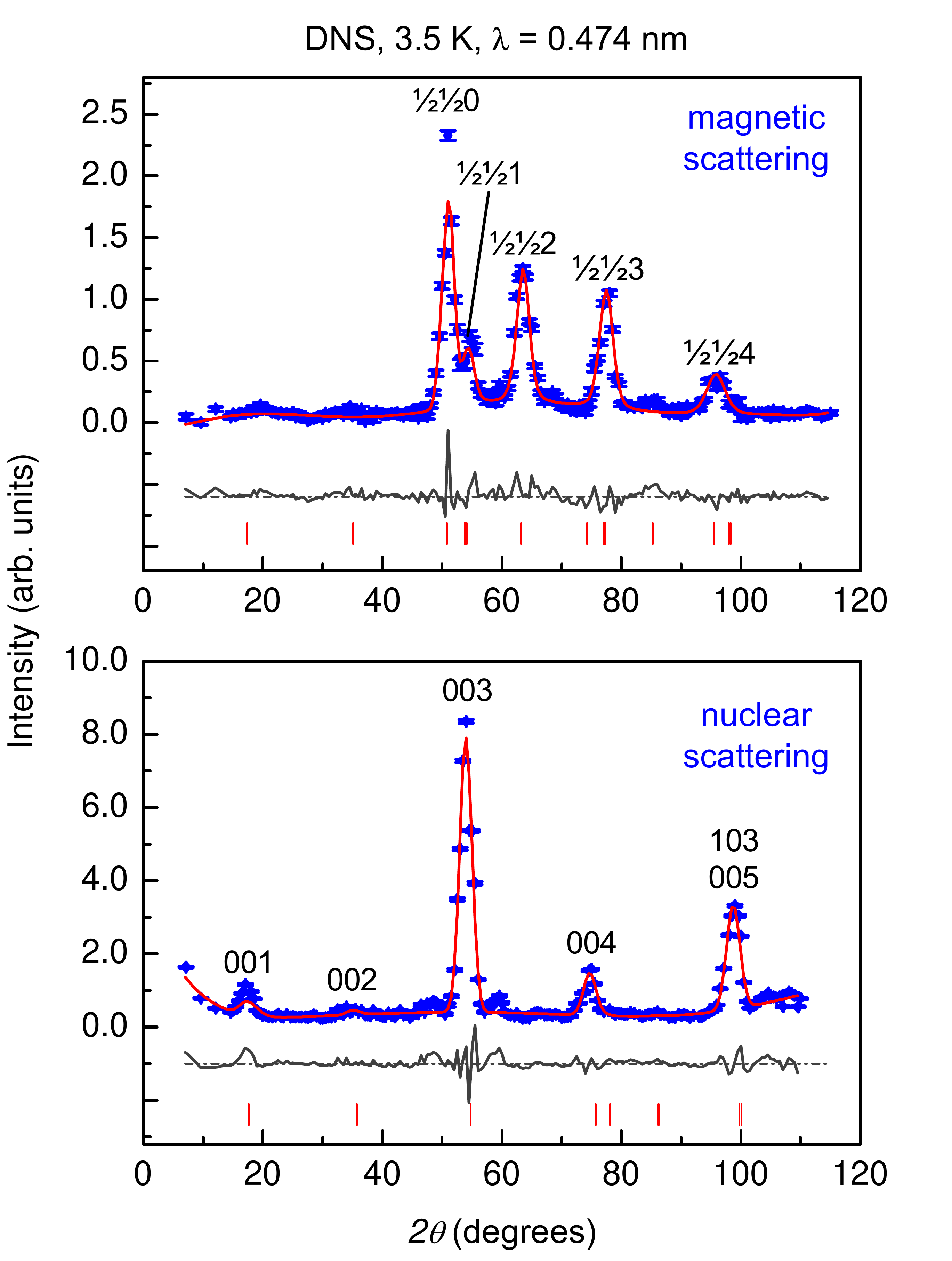}
\caption{(Color online) Magnetic and nuclear reflections of \SCOFA~(blue) and Rietveld fit (red) at 3.5~K measured at the polarized spectrometer DNS. The magnetic space group is \MSG~and the crystallographic space group \CSG. The miller indices of the magnetic reflections were transformed to the tetragonal cell for comparability. Red markers: Reflection conditions of both space groups.}
\label{fig:Rietveld_dns}
}
\end{figure}

\begin{figure}[h!]
\center{
\includegraphics[width=65mm,clip]{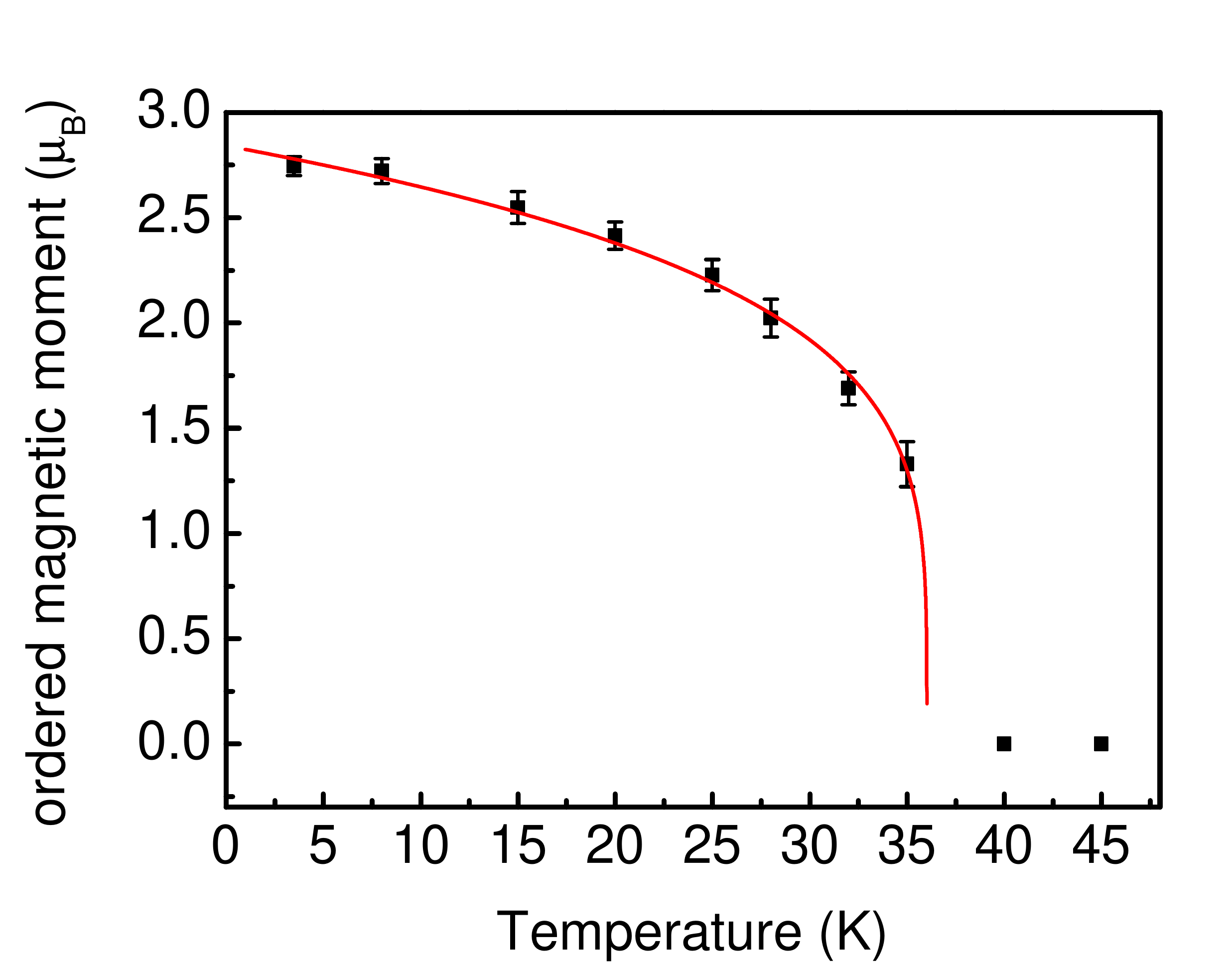}
\caption{Variation of the refined ordered magnetic moment of the Cr-sublattice with temperature. The temperature dependence follows the simple power law \PLAW~(red curve).}
\label{fig:mu}
}
\end{figure}

From the observed $\mathbf{q}$-vector $(\frac{1}{2},\frac{1}{2},0)$ we assumed a checkerboard-like spin arrangement of the $C$-type, which is reversed between the adjacent chromium layers along $\mathbf{c}$. Since the layers are at the coordinates $z = \pm0.31$, no $G$-type pattern is possible. A first expected spin alignment along $\mathbf{c}$ did not reproduce the observed data, therefore we developed models with orientations within the $(\mathbf{ab})$-plane. The by far best fit was found with the alignment along $[\mathbf{a-b}]$ and reversed along $\mathbf{a}$ and $\mathbf{b}$ based on the original tetragonal cell. This arrangement required the orthorhombic magnetic space group \MSG~(Litvin No. $67.15.591$)\cite{litvin}, where \CR~occupies the $4g$ ($0,\frac{1}{4},z$) Wyckoff position and the magnetic moments of \CR~align along the orthorhombic $\mathbf{a'}$-axis, building up a checkerboard arrangement in each Cr layer at both heights $z$ and $\overline{z}$ (Figure~\ref{fig:order}). 
By testing different magnetic space groups and different spin orientations, any other model for the magnetic ordering in a direct magnetic subgroup derived from the crystallographic space group $Cmme$ could be unambiguously ruled out. Sections of the nuclear and magnetic powder patterns recorded at 3.5~K and corresponding Rietveld refinements are depicted in Figure~\ref{fig:Rietveld_dns}. The ordered magnetic moment of a \CR-ion was refined to be $2.75(5)~\mu_B$ at 3.5~K, which is close to the expected $3~\mu_B$. The evolution of the magnetic ordering becomes evident from the order parameter and its temperature dependence is depicted in Figure~\ref{fig:mu}. It follows the power law \PLAW~with $a=2.84(3)$, $T_N=36.0(5)$~K and $\beta=0.22(2)$. The exponent $\beta$ is between the idealized 2D and 3D Ising values ($\frac{1}{8}$ and $\frac{5}{16}$, respectively), which agrees with the fact that the magnetic arrangement of each Cr-layer at height $z$ is coupled with the corresponding $\overline{z}$-layer and the magnetism therefore can be explained neither strictly two- nor three-dimensionally.


\section{Conclusion}
We have synthesized the iron arsenide oxide \SCOFA~and re-determined its crystallographic and magnetic structure by neutron diffraction experiments at the D20 and DNS diffractometers. A mixed occupancy of chromium and iron in the FeAs layers was found, which points up the ability of substitution between similar $3d$-metals in these compounds. We suggest that this Cr-doping may also be the reason for the absence of a SDW anomaly and superconductivity. Such non-stoichiometries may also occur in similar compounds like the superconducting Sr$_2$VO$_3$FeAs and may also be responsible for the different physical behavior of these compounds when compared with the 1111- and 122-iron arsenides. Deviations from the ideal stoichiometry has especially to be taken into account when discussing their electronic structures. \SCOFA~shows short-range spin correlations from the \CR-ions below $\sim 100$~K and long-range antiferromagnetic ordering below $T_N = 36.0(5)~K$. The magnetic structure is of the $C$-type with the Cr-spins oriented parallel to $[\mathbf{a-b}]$ with all nearest-neighbor \CR~moments antiferromagnetically aligned, thus forming a checkerboard arrangement.

\acknowledgments
M. T. would like to thank Dr. Klaus R\"{o}mer for financial support. This work was financially supported by the German Research Foundation (DFG).

\end{document}